\documentclass[prb,superscriptaddress,showpacs,twocolumn]{revtex4-1}
\makeindex
\usepackage{amsmath}
\usepackage{amsthm}
\usepackage{amssymb}
\usepackage{mathrsfs}
\usepackage{float}
\usepackage[pdftex]{graphicx}
\usepackage{xcolor}
\definecolor{link}{rgb}{0.8, 0.8, 0.8}
\usepackage[linkbordercolor=link]{hyperref}
\usepackage{calc}
\providecommand{\includegraphics}[2][width=\textwidth]{$#2$}

\usepackage{supertabular}
\usepackage{comment}

\def\figscale{0.4}
\def\figscalebis{0.46}

\def\RR{\mathbb R}

\def\E{{\mathcal E}}

\def\XX{\overline{X}}
\def\YY{\overline{Y}}
\def\ra{\textsf{R}_a}
\def\seff{\sigma_{\mathrm{eff}}}
\def\Q{\mathrm{Q}}

\begin{document}

\title{Confidence and efficiency scaling in Variational Quantum Monte Carlo calculations}
\author{F. Delyon}
\affiliation{LPTMC, UMR 7600 of CNRS, Universit\'e P. et M. Curie, Paris, France}
\author{B. Bernu}
\affiliation{LPTMC, UMR 7600 of CNRS, Universit\'e P. et M. Curie, Paris, France}
\author{Markus Holzmann}
\affiliation{LPMMC, UMR 5493 of CNRS, Universit\'e Grenoble Alpes, B.P. 166, 38042 Grenoble, France}

\date{\today}

\begin{abstract}
Based on the central limit theorem, we discuss the problem of evaluation of the statistical error of Monte Carlo
calculations using a time discretized diffusion process. 
We present a robust and practical method to determine the effective variance of general observables
and show how to verify the equilibrium hypothesis by the Kolmogorov-Smirnov test. 
We then derive scaling laws of the efficiency illustrated by Variational Monte Carlo calculations on the two dimensional electron gas.
\end{abstract}
\maketitle

\section{Introduction}

Monte Carlo integration techniques have become a standard tool in statistical physics and classical and quantum many body theory \cite{Binder,Kalos}. However, due to the finite simulation time, outcomes of such calculations are affected by statistical uncertainties and the
precise estimation of the resulting error is essential to obtain quantitative results. 

Standard Monte Carlo methods are based on Markov diffusion processes using Metropolis-Hastings algorithm. There, sequential outputs 
are not independent, and, similar to any other statistical method, the main problem is to estimate the robustness of the sampling.

 As one usually deals with a very large configuration space, an exact answer is impossible. Modestly, one expects that a finite number of samples may reflect the expectation on the whole space, and we want to ensure the coherence of our statistics. 
That is, we want to check the stationarity  of our sampling and to estimate the accuracy of some averaged quantity.

The aim of this work is to show that the central limit  theorem (CLT) and the Kolmogorov-Smirnov theorem provide simple tools  to 
determine the accuracy of some observable and to test the coherence of the sampling.
These well known tools have already been  discussed in this context\cite{Binder,Hastings,Caflish}, here we provide an effective implementation of these mathematical results.

In the following, most examples come from Variational Quantum Monte-Carlo (VMC) calculations
applied to the 2D homogeneous electron gas in a quadratic box of length $L$ with periodic boundary conditions\cite{Vignale}.
The configuration space is $[-L/2,L/2]^{\nu}$ where $\nu= 2 N_e$ for  the $N_e$ electrons, and we define
the dimensionless parameter $r_s$ by 
$\pi a_B^2r_s^2 N_e=L^2$ where $a_B$ is the Bohr radius.

Given a complex antisymmetric function $\Psi(r_1,\ldots ,r_{N_e})$ of the Slater-Jastrow form, 
we concentrate on one of the most important quantity, the average energy, $\E$, of this state:
\begin{align}
\E=\frac{\langle \Psi|H|\Psi\rangle}{\langle \Psi|\Psi\rangle}
\end{align}
where $H$ is the electronic Hamiltonian.

In fact the above integration is rapidly unfeasible as the number of electrons increases. In Sect.II we briefly explain
the standard VMC approach to compute the above integral for large values of $N_e$. Our estimation of the statistical error
relies on the CLT is described and tested in Sect.III. In Sect. IV, we show how the Kolmogorov-Smirnov test can be implement
to decide on the consistency of the statistical distributions. The efficiency and optimal behavior of VMC with 
 respect to the size $N_e$ and the discretization of the Markov process is discussed in Sect.V.

\section{Discretized diffusion algorithm}\label{VMC}
We start with a brief description of the standard algorithm used for Variational Monte Carlo calculations.

Let $P(R)dR$ be a probability on $\RR^\nu$,
the idea of Monte Carlo methods \cite{Kalos} is  to calculate expectations using a multidimensional diffusion process.
Indeed, if $R(t)$ is an ergodic Markov process with invariant measure $P(R)dR$ then:
\begin{align}
\lim_{T\rightarrow \infty} \frac 1T\int_0^Tf(R(t))dt=\int f(R) P(R)dR
\end{align}
provided that $\int |f(R)| P(R)dR<+\infty$.

Thus, one has to choose a  diffusion process such that the invariant distribution is exactly $P$.
Let $R(t)$ be given by the Langevin equation\cite{Brei}:
 \begin{align}
 \label{wiener}
dR(t)=G(R)dt +dw
\end{align}
$w$ is a vector of $\nu$ independent Brownian motions (Wiener processes) and $G$ a vector of $\nu$ regular functions.

The corresponding Kolmogorov\cite{Kolmo} (or Fokker-Planck) forward equation for a measure $\mu(R,t)dR$ is:
 \begin{align}
 \label{timeev}
\frac{\partial \mu}{\partial t}=-\nabla G\mu+\frac 12\Delta \mu
\end{align}
and choosing 
\begin{align}
G=\frac 12\nabla \ln (P)
\end{align}
$P(R)dR$ is an invariant measure of Eq.~(\ref{timeev}). 

For numerical simulations, one approximates the process, Eq.~(\ref{wiener}), with the following discrete process:
 \begin{align}
 \label{wienerdisc}
R_{i+1}=R_i+G(R_i)\tau +  w(\tau)
\end{align}
As $\tau$ goes to zero we expect that the invariant probability of  Eq.~(\ref{wienerdisc}) goes to $P$.
The integral kernel  of the Markov process (\ref{wienerdisc}) is:
\begin{align}
 \label{kernel}
T(R'|R)=\frac 1{(2\pi\tau)^{\nu/2}}\exp(-\frac{(R'-R-G(R)\tau)^2}{2\tau})
\end{align}
In practice, to avoid convergence analysis, one uses the Metropolis\cite{Metro} algorithm to obtain precisely $P$ as the invariant measure.
 First, starting at $R$, we choose the next point $R'$ according to Eq.~( \ref{kernel}).
 Thereafter, we use an auxiliary boolean independent variable in order to accept or reject the new point with probability $A(R',R)$ such that $P$ is invariant. Equivalently, we have to choose $A(R',R)$ such that the mean value of any function $f(R)$ is invariant
\begin{align}
\nonumber
	\int f(R)P(R)dR=&\int f(R') A(R',R)T(R'|R)P(R)dRdR'\\
\nonumber
	+ \int f(R)P(R) & \left[ 1- \int A(R',R)T(R'|R) dR' \right] dR
\end{align}
i.e.
\begin{align}
\nonumber
	\int \!\!dR'\,T(R|R')&A(R,R')P(R')=\\
		&\int \!\!dR'\,T(R'|R)A(R',R)P(R).
\end{align}
This condition is fulfilled imposing detailed balance:
\begin{align}
T(R|R')A(R,R')P(R')=T(R'|R)A(R',R)P(R).
\end{align}
Hence, the optimal (minimal rejection) solution is given by
\begin{align}
\label{Q}
A(R',R)=\min\left(1,\frac {T(R|R')P(R')}{T(R'|R)P(R)}\right).
\end{align}
If we reject the new point, the point $R$ is counted twice.
Starting from $R_1$ and iterating this process, we obtain a sequence of vectors $(R_1,R_2,\ldots)$ and a sequence of integral weights $(n_1,n_2,\dots)$ where
$n_i$ is the number of repetitions of the vector $R_i$. This sequence asymptotically reflect the distribution $P$.

The acceptance rate $\ra$, is the probability that a move is accepted, thus
\begin{align}
\ra=\frac{\sum 1}{\sum n_i}\rightarrow\frac{1}{E(n)}
\end{align}
where $E(.)$ stands for the expectation of a random variable.

For a function $X(R)$, we associate the process $(X_i=X(R_i))$, and the empirical expectation of the variable $X$ is given by:
\begin{align}\label{mean}
\XX_N=\frac{\sum_{i=1}^{N} X_in_i}{\sum_{i=1}^{N} n_i}
\end{align}
The main problem is to estimate the robustness of a sampling $\{X_i,n_i\}$.

In our case, $\nu=2N_e$ and $R$ is sampled with a probability proportional to $|\Psi(R)|^2dR$; that is $G(R)=\nabla \ln|\Psi(R)|$.

Setting
\begin{align}
X_i(R_i)=\frac{(H\Psi)(R_i)}{\Psi(R_i)}
\end{align}
we expect for a fair sampling of the total energy
\begin{align}
\frac {\sum_i X_in_i}{\sum_i n_i}\rightarrow \frac {\int (H\Psi)(R)\overline{\Psi}(R)dR}{\int |\Psi(R)|^2dR}=\frac{\langle \Psi|H|\Psi\rangle}{\langle \Psi|\Psi\rangle}
\end{align}

\section{The central limit theorem}
The central limit theorem(CLT) is a well-known  result for independent identically distributed random variables providing the expected 
fluctuations of the mean of a random sequence.
This theorem has many generalizations in particular for Markov process\cite{Kipnis,Geyer,Bill,Jones}.

Let $(Y_1, Y_2,\ldots)$ be a stationary mixing Markov process with invariant probability $\pi$ and $E(|Y|)<+\infty$.
Let 
\begin{align}
	\tilde Y_i=Y_i-E(Y)
\end{align}
and ${\mathcal N}(a,\sigma^2)$ be  the Gaussian law of mean $a$ and variance $\sigma^2$, then the ergodic theorem guarantees that
\begin{align}
Z_N=\frac 1{\sqrt N}\sum_{i=1}^N\tilde Y_i 
\end{align}
goes to zero with probability one.
A central limit theorem (CLT) for Markov process gives conditions under which 
\begin{align}
Z_N\xrightarrow{\ {\mathcal D}\ }  {\mathcal N}(0,\seff^2)
\end{align}
where
\begin{align}
\label{sigma}
\seff^2=E(\tilde Y_1^2)+2\sum_{k>1}E(\tilde Y_1\tilde Y_k).
\end{align}
and ${\mathcal D}$ stands for the convergence in distribution.
Equation~(\ref{sigma})  can be easily guessed since $\seff^2$ corresponds to the limit of $E\left(Z_N^2\right)$
as soon as 
\begin{align}
\label{Ckdef}
C_k=E(\tilde Y_1\tilde Y_{k+1})
\end{align}
decreases rapidly as $k$ increases.
Here we suppose that this is our case and that the CLT is in force; the problem is how to estimate  $\seff$ for finite samplings.
     
\subsection{Determination of the effective variance}
For an empirical sampling $(Y_i)_{i=1}^N$, the empirical estimate for $E(Y)$ is
\begin{align}
\label{defY}
\YY_N=\frac 1{ N} \sum_{i=1}^NY_i
\end{align}
and if we define
\begin{align}
C_{N,k}=\frac 1{ N} \sum_{i=1}^{N-k}(Y_i-\YY_N)(Y_{i+k}-\YY_N)
\end{align}
we have $\lim_{N\to\infty} C_{N,k}=C_k$. We then have
\begin{align}
\label{sigma3}
\seff^2&=\lim_{N\rightarrow \infty}C_{N,0}+2\lim_{n\rightarrow \infty}\sum_{k=1}^n\lim_{N\rightarrow \infty}C_{N,k}
\end{align}
Notice that the order of the two limits is important, as we also have
\begin{align}
\nonumber
C_{N,0}+2\sum_{k}C_{N,k}&=\frac 1{ N} \sum_{i,j}(Y_i-\YY_N)(Y_{j}-\YY_N)\\
&=0
\end{align}
In order to impose the right order of the limit, we use a so-called window estimator \cite{Geyer} in the following
 \begin{align}
\seff^2(k)=C_{N,0}+2\sum_{i=1}^{k}C_{N,i}
\end{align}
with $1 \ll k \ll N$. The main difficulty is the choice of $k$ such that $\seff^2(k_l)$ provides a robust estimate
for the true uncertainty: too small values of $k$ may considerably underestimate the error, whereas large values of $k_m$ will
mainly add noise such that $\seff^2(k)$ becomes unreliable.

Indeed, even if we assume that $C_{k}$ decreases quickly,
 for large $k$, $C_{N,k}$ is a random variable of order $1/\sqrt N$. From
\begin{align}
\nonumber
C_{N,k}=&\frac 1{ N} \sum_{i=1}^{N-k}(\tilde Y_i+E(Y)-\YY_N)(\tilde Y_{i+k}+E(Y)-\YY_N)\\
\nonumber
=&\frac 1N\sum_{i=1}^{N-k}\tilde Y_i\tilde Y_{k+i}-(E(Y)-\YY_N)^2\\
&-\frac 1N\left(\sum_{i=1}^{k-1}\tilde Y_{i}+\sum_{i=N-k}^N\tilde Y_{i}\right)(E(Y)-\YY_N)\\
\label{Cstart}
=&\left(\frac 1N\sum_{i}\tilde Y_i\tilde Y_{k+i}\right)-O(1/N),
\end{align}
we get
\begin{align}
NC_{N,k}^2=\frac 1N\sum_{i,j}\tilde Y_i\tilde Y_{k+i}\tilde Y_j\tilde Y_{k+j}-C_kO(1).
\end{align}
and since $C_k$ is assumed to vanish, we get 
\begin{align}
\label{C2}
	E(NC_{N,k}^2)\rightarrow E(Y_0^2)+2\sum_{j=1}^{+\infty} E(\tilde Y_0\tilde Y_{k}\tilde Y_j\tilde Y_{k+j}).
\end{align}
For large $k$, $\tilde Y_0$ (resp. $\tilde Y_j$) and $\tilde Y_{k}$ (resp.$\tilde Y_{k+j}$) are centered  independent variables, therefore  the non-zero terms in Eq.~(\ref{C2}) are for small $j$, and evaluate as:
\begin{align}
\label{C3}
	E(\tilde Y_0\tilde Y_{k}\tilde Y_j\tilde Y_{k+j})\xrightarrow {k\rightarrow\infty} E(\tilde Y_0\tilde Y_j)E(\tilde Y_{k}\tilde Y_{k+j})=C_{j}^2
 \end{align}
Therefore, Eq.~(\ref{C2}) asymptotically becomes
\begin{align}
\label{C4}
	\Sigma^2=\lim_{k\rightarrow \infty}\lim_{N\rightarrow \infty}NE(C_{N,k}^2)= C_0^2+2\sum_{j=1}^{+\infty} C_{j}^2
\end{align}
leading to fluctuations of order $N^{-1/2}$ for $C_{N,k}$ at large $k$.

{\bf Empirical estimation of $\seff$}, Eq.~(\ref{sigma3}).
 Let us set:
 \begin{align}
 \label{bigsigma}
\Sigma_{N,k}^2=C_{N,0}^2+2\sum_{j=1}^{k} C_{N,j}^2.
\end{align}
$\Sigma_{N,k}^2$ is an increasing function of $k$ and for large $k$ it goes like $\Sigma^2(1+k/N)$.
On the other hand $C_{N,k}^2$ is of order one for small $k$ then decreases and oscillates around $\Sigma^2/ N$.
Thus there is a marginal value $k_m$ corresponding to the first $k$ such that  $NC_{N,k}^2<\Sigma_{N,k}^2$ providing 
a good estimate of the effective variance, Eq.~(\ref{sigma3}).
Clearly, this estimate makes sense only if $\sqrt NC_{N,k_m}\ll C_0$, otherwise we must consider that $N$ is not large enough.

In the following subsections, we demonstrate the robustness of our error estimation
and test its validity against independent data sets.

\subsection{CLT and Monte Carlo}
We can extend the CLT to Monte-Carlo observables considering the weight $n_i$  of the Metropolis part (see section \ref{VMC}).
Now $\XX_N$ is given by Eq.~(\ref{mean}) and setting $\tilde X_i=X_i -E(X)$
\begin{align}
N\left(\frac {\sum \tilde X_i n_i}{\sum  n_i}\right)^2\rightarrow C_{N,0}+2\sum_k C_{N,k}
\end{align}
where:
\begin{align}
\label{corrN}
C_{N,k}&=\frac 1{ E(n)(\sum_i n_i)} \sum_{i=1}^{N-k} n_i\tilde X_in_{i+k}\tilde X_{i+k}
\end{align}
and Eq.~(\ref{C4}) is still in force.

\begin{figure}[H]
\begin{center}
\includegraphics[width=\figscalebis\textwidth]{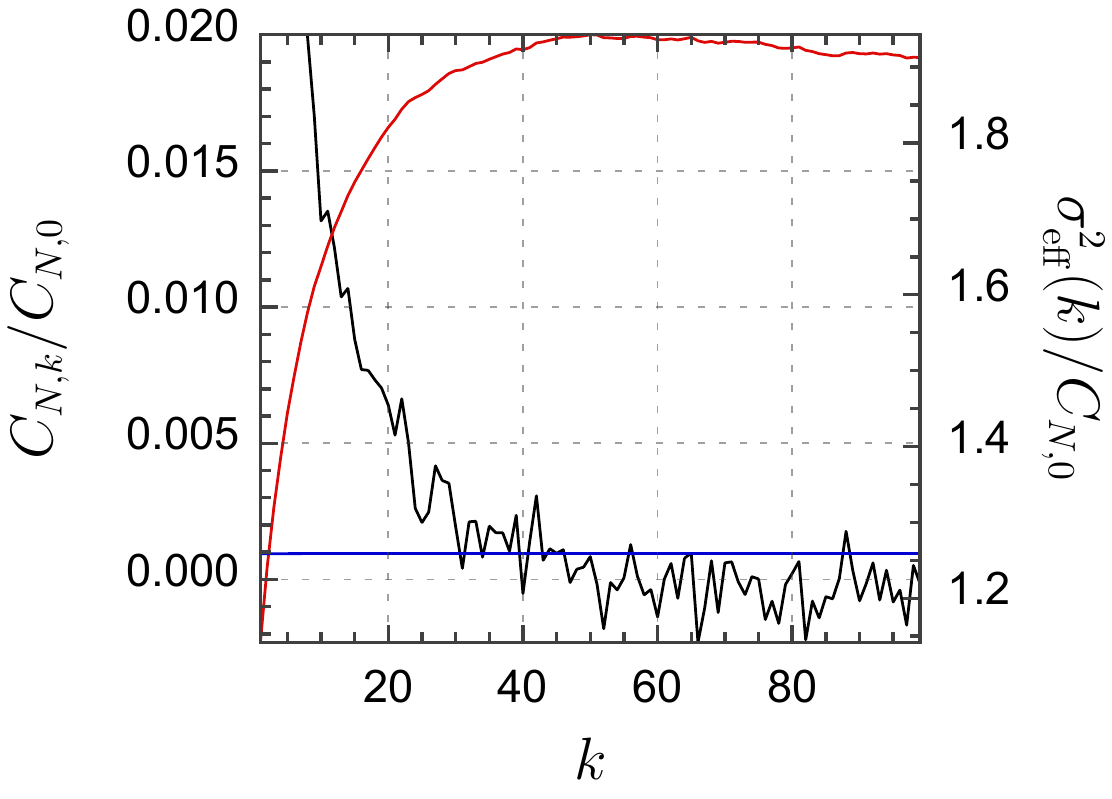}
\caption{
$C_{N,k}/C_{N,0}$ as a function of $k$ (black line) and $\seff^2(k)/C_{N,0}$ (red line). The blue line is 
$\Sigma_{N,k}/(\sqrt{N}C_{N,0})$;
The crossing of the black and blue curves occurs for  $k=31$ giving $\seff^2=1.90\ C_0$.
}
\label{FIG-Correlation}
\end{center}
\end{figure}

\begin{figure}[H]
\begin{center}
\includegraphics[width=\figscale\textwidth]{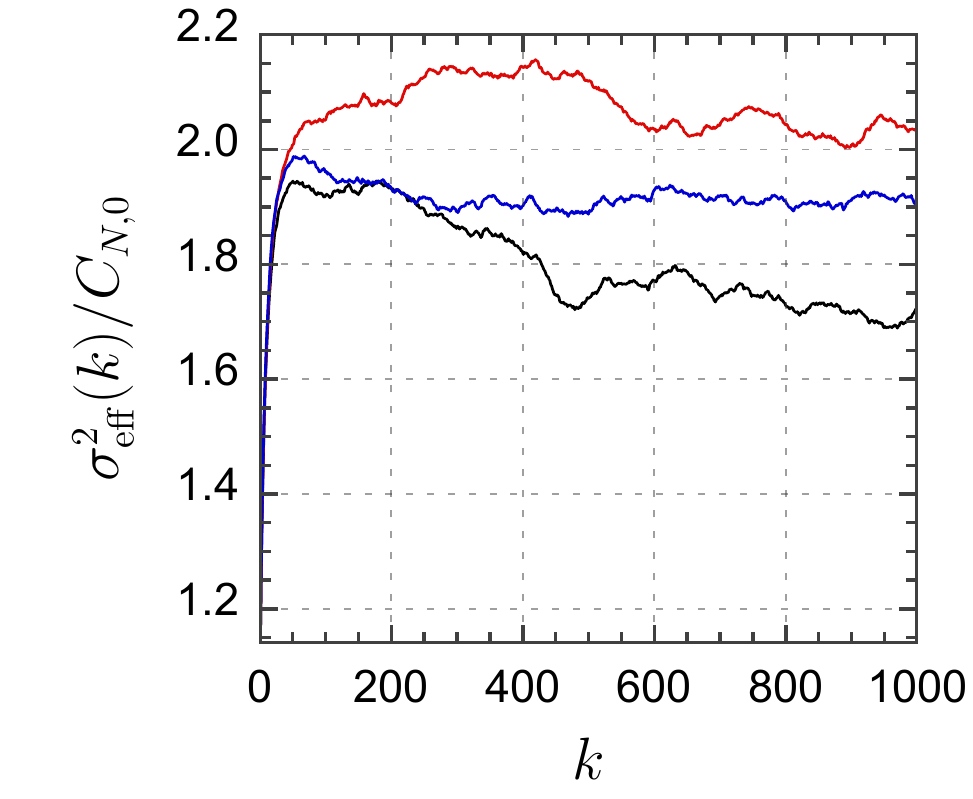}
\caption{$\seff^2(k)/C_0$ for three independent samplings.
}
\label{FIG-Correlations}
\end{center}
\end{figure}

Fig.~\ref{FIG-Correlation} shows a representative example of the behavior of $C_{N,k}$ for a MC simulation with an acceptance rate $\ra$ of 0.45.
Furthermore, Eq.~(\ref{C3}) may be extended to estimate the correlations of $C_{N,k}$:
\begin{align}
E(NC_{N,k'}C_{N,k+k'})\xrightarrow {k'\rightarrow\infty} \sum_{i=-\infty} ^\infty C_i C_{i+k}
\end{align}
This shows that $C_{N,k}$ may be strongly correlated.
Looking at the red curve of Fig.~\ref{FIG-Correlation}, the cutoff at $k=31$ may seem unjustified, but different samplings 
lead to different behavior after this cutoff while the behavior for smaller values of $k$ is robust (Fig.~\ref{FIG-Correlations}). 

\subsection{Comparison with sample fluctuations}
We can compare our estimate of the empirical variance with the variance obtained with  independent samples.
We make $p=40$ independent samples of $N$=45000 trials.

For each sample $\alpha$ we compute the mean:
\begin{align}
\overline X_\alpha=\frac {\sum_{i=1}^N X_{\alpha,i}n_{\alpha,i}}{ \sum_{i=1}^N n_{\alpha,i}}
\end{align}
and we get the empirical mean:
\begin{align}
\overline X=\frac 1 {p} \sum_{\alpha=1}^p \overline X_\alpha
\end{align}
The variance of $X$ (corresponding to $C_{N,0}$ in Eq.~(\ref{corrN})) is 0.308 and taking into account  the correlations, Eq.~(\ref{sigma}), we get $\seff^2$ is  0.591.
This gives a standard deviation $\frac \seff{\sqrt N}$=0.00363 for the $X_\alpha$'s to be compared with 0.00370 obtained directly from the variance of  the 40 values of  $\overline X_\alpha$. 

\begin{figure}[H]
\begin{center}
\includegraphics[width=\figscale\textwidth]{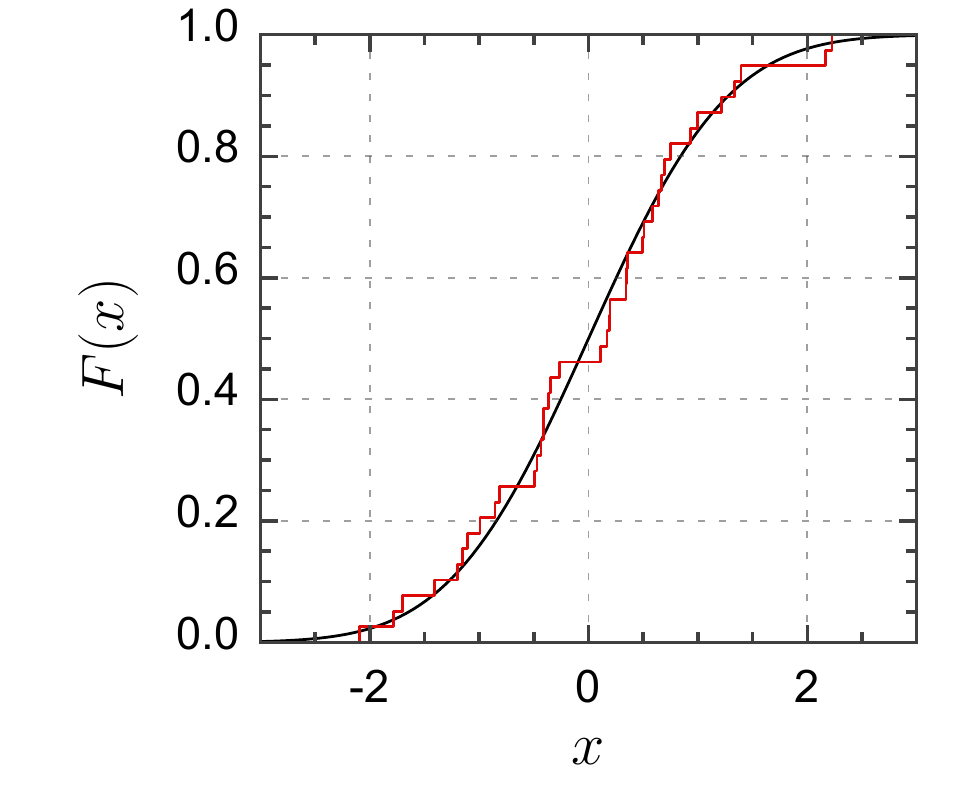}
\caption{Distributions of $x=\frac{\overline  X_\alpha-\overline  X}\seff$ for 40 independent samples of 45000  trials. The black line is the distribution function 
of the normalized Gaussian.}
\label{GaussianTest}
\end{center}
\end{figure}

Now we can go further and check the asymptotic normal law of the $X_\alpha$'s. In Fig.~\ref{GaussianTest}, we plot the distribution of:
\begin{align}
\label{xalpha}
x_\alpha=\sqrt{N}\frac{\overline X_\alpha- \overline X}\seff.
\end{align}
supposed to be a normalized centered Gaussian variable.
The black curve is the distribution function of a normalized Gaussian.
Since the $\overline X_\alpha$ are independent, we have:
\begin{align}
\label{corrx}
M_{\alpha \beta}=E(x_\alpha x_\beta)\xrightarrow [N\rightarrow \infty]{}-\frac 1p+\delta_{\alpha \beta}
\end{align}
The correlation matrix $M$ has eigenvalues 0 (multiplicity 1) and 1 (multiplicity $p-1$).
Thus for large $N$, the law of random variable  $S=\sum_{\alpha=1}^p x_\alpha^2$ is exactly the law of the sum of the square of $p-1$ normalized Gaussian variables.
Therefore, we can use the $\chi^2$ test:
\begin{align}
	P( S<x)=\chi^2_{p-1}( x)={\mathrm \Gamma}\left(\frac{p-1}2, \frac x2\right)
\end{align}
where ${\mathrm \Gamma}$ is the regularized incomplete gamma function.
Here we have $S=40.1$ and the $\chi^2_{39}( 40.1)=0.62$.

\begin{figure}[H]
\begin{center}
\includegraphics[width=\figscalebis\textwidth]{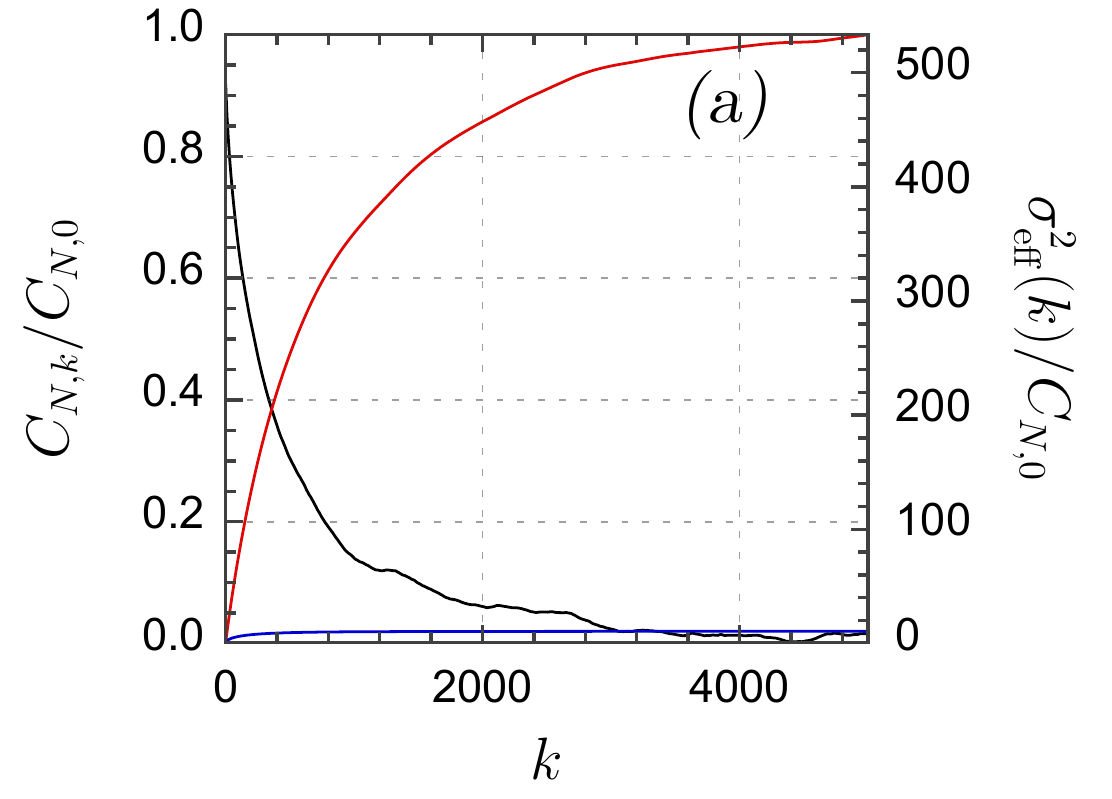}
\includegraphics[width=\figscale\textwidth]{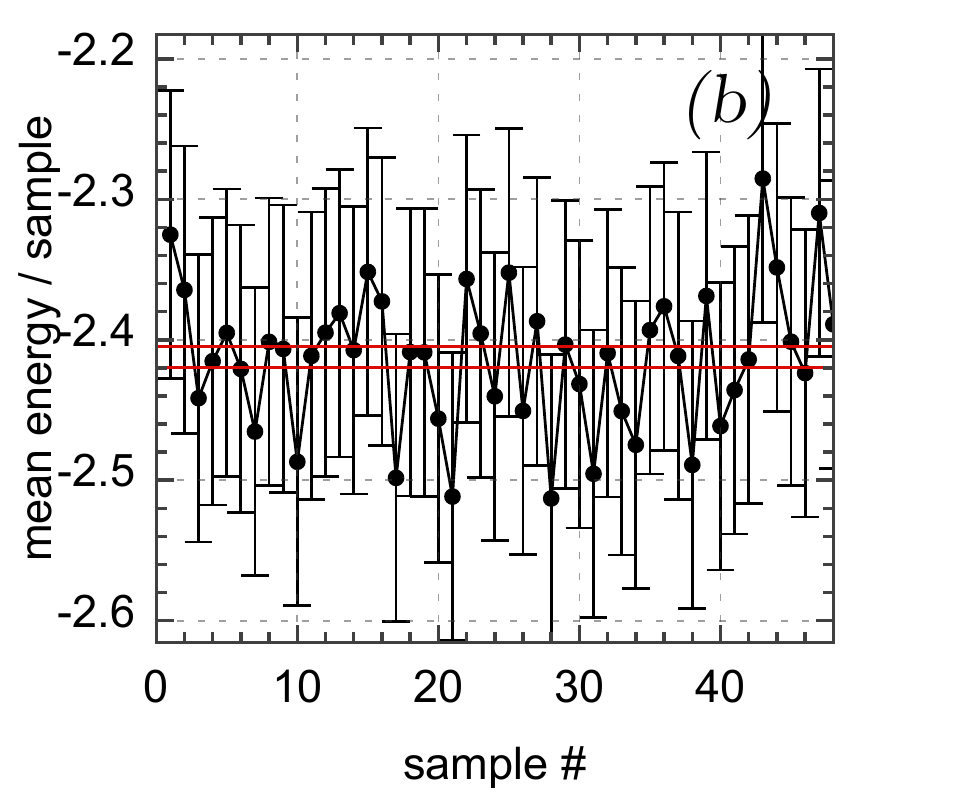}
\caption{Example of long correlated walks ($\ra=0.94$). (a) Correlations (black line) and running sum of correlations (red line). The blue line is $\Sigma_{N,k}/(\sqrt{N}C_{N,0})$, Eq.~(\ref{bigsigma}). (b) mean energy for the 48 samples (arbitrary units).}
\label{Emean}
\end{center}
\end{figure}

As the acceptance rate $\ra$ approaches 1, the dynamic is very slow and the correlations 
are very important leading to large fluctuations.
The following sampling is made of 48 samples of 19000 records. The acceptance rate $\ra$ is 0.94. In Fig.~\ref{Emean} the scaled correlations $C_k$ are relevant until $k=3244$ leading to $\sum_{k>0} C_k=530C_0$. Thus $\seff^2$ is about 1060 times larger than $C_0$ leading to an accuracy of 0.00724.
This increase of the fluctuations is well verified by the distribution on the 48 means giving an accuracy of 0.00728. The $\chi^2$ test gives 0.53.

However, in this case $\seff^2$ is computed with the 48 samples leading to a marginal $k_m$ such that $C_{k_m}/C_0\approx 0.02$. The estimate for  only one sample 
gives 0.2 which cannot be consider as small. Thus the 48 samples are relevant for the CLT while a single sample cannot provide any estimate of the accuracy.

\section{The Kolmogorov-Smirnov test}\label{KSsection}
Here, we want to provide a quantitative
test, based on the Kolmogorov-Smirnov theorem, to verify if the samplings are actually consistent with an equilibrium hypothesis.
\subsection{The Kolmogorov-Smirnov theorem}
First, we briefly recall here the definition and the theorem\cite{Brei,Kolmo}.
Let $(Y_1,Y_2,\ldots)$ be independent, identically distributed random variables with continuous distribution function $F$.
Let $F_N(x)$ be the empirical distribution function of $(Y_1,Y_2,\ldots Y_N)$:
\begin{align}\label{fn0}
F_N(x)=\frac 1N\sum_i \chi(Y_i<x)
\end{align}
where  $\chi(b)=1$ if $b$ is true otherwise 0.
As $N$ goes to infinity, by the law of large numbers, $F_N(x)\rightarrow F(x)$ almost surely. 
From the definition: 
\begin{align}\Delta_N(x)=\sqrt{N}(F_N(x)-F(x)), 
\end{align}
we have $E(\Delta_N(x))=0$ and 
a straightforward calculation gives for $x\le y$:
\begin{align}\label{fn3}
E(\Delta_N(x)\Delta_N(y))&=F(x)(1-F(y))
\end{align}
Thus the CLT for independent variables guarantees
\begin{align}\label{delta0}
\Delta_N(x)\xrightarrow {\ {\mathcal D}\ }{\mathcal N}\left(0,F(x)(1-F(x))\right)
\end{align}
Let $B(t)$ be the normalized Brownian bridge: the law of Brownian bridge is law of a Brownian motion $b(t)$ such that $b(1)=0$
(equivalently the law of Brownian bridge $B(t)$ is the law of $b(t)-tb(1)$). For $t\le t'$
\begin{align}
E(B(t)B(t'))&=t(1-t')
\end{align}
thus $E(\Delta_N(x)\Delta_N(y))$ is exactly the correlation of the Brownian bridge $E(B(t)B(t'))$ at times $t=F(x)$ and $t'=F(y)$.

More precisely, the Kolmogorov-Smirnov  theorem tells that
\begin{align}
\label{KS}
D_N=\max_x|\Delta_N(x)|\xrightarrow{\ {\mathcal D}\ }D= \max_{0\le t\le 1} |B(t)|.
\end{align}

\begin{figure}[h]
\begin{center}
\includegraphics[width=\figscale\textwidth]{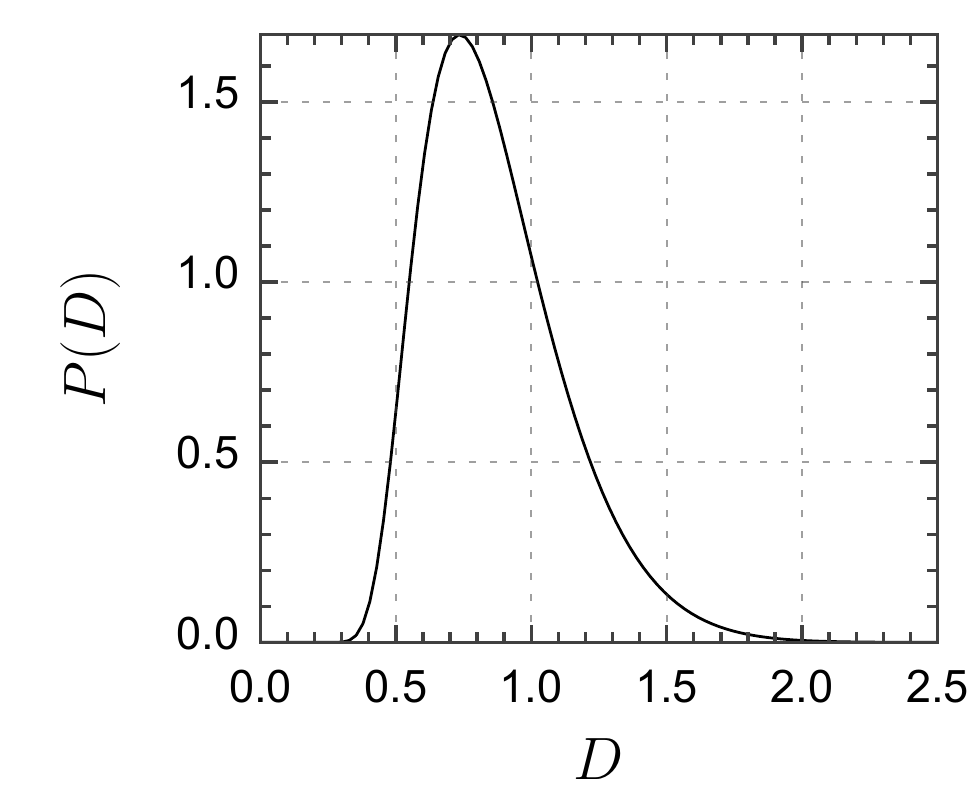}
\caption{Density of $D=\max_{0\leq t\leq 1} |B(t)|$}
\label{FIG-Comparison}
\end{center}
\end{figure}

The important point is that the law of $D$ does not depend on the distribution $F$ providing the non-parametric  K-S test\cite{KS}.
The Kolmogorov-Smirnov test evaluates the probability $P(D>D_N)$. The law of  $D$ is given by:
\begin{align}
\label{PBB}
 P(D<x)&=1+2\sum_{n\ge 1}(-1)^ne^{-2n^2x^2}\\
&=\frac{\sqrt{2\pi}}{x}\sum_{n\ge 0}e^{-(2n+1)^2\pi^2/(8x^2)}
\end{align}
The density of the probability of $D$ is given on Fig.~\ref{FIG-Comparison}
and for instance $P(D>2)\approx 7.\,10^{-4}$ and $P(D<0.3)\approx 9.\,10^{-6}$.

This is the viewpoint of statistics as you need to choose the best candidate among a family of known distribution functions $F_\lambda$. 
Let us now consider the case where the distribution $F(y)$ in unknown.
\subsection{Implementation for unknown distribution.}
A simple approach is to divide your sample into $p$ samples  $\{Y_\alpha\}_{\alpha=1\ldots p}$ of length $N$.
In order to check  homogeneity of the sampling, we first build the distribution $\hat F(y)$ of all the samplings.
Thereafter, for each samples $Y_\alpha$ we build the distribution function $F_\alpha(y)$ and the differences
\begin{align}
\label{deltaa}
\Delta_\alpha(y)=\sqrt N(F_\alpha(y)-\hat F(y)).
\end{align}
One checks that for $x<y$:
\begin{align}
 E(\Delta_\alpha(x)\Delta_\beta(y))&= M_{\alpha\beta}F(x)(1-F(y))\\
 M_{\alpha\beta}&=\delta_{\alpha\beta}-\frac 1p
\end{align}
Thus the $p$ processes are not rigorously independent. As above, the correlation matrix $M$ has $p-1$ eigenvalues equal to $1$ and
 a null eigenvalue corresponding to the vector $(1,1,\ldots,1)$. Thus, they represent $p-1$ independent Brownian bridges and the distribution of 
$D_\alpha=\max _{y}\Delta_\alpha(y)$ must be close to distribution of $D$. 

\begin{figure}[h]
\begin{center}
\includegraphics[width=\figscale\textwidth]{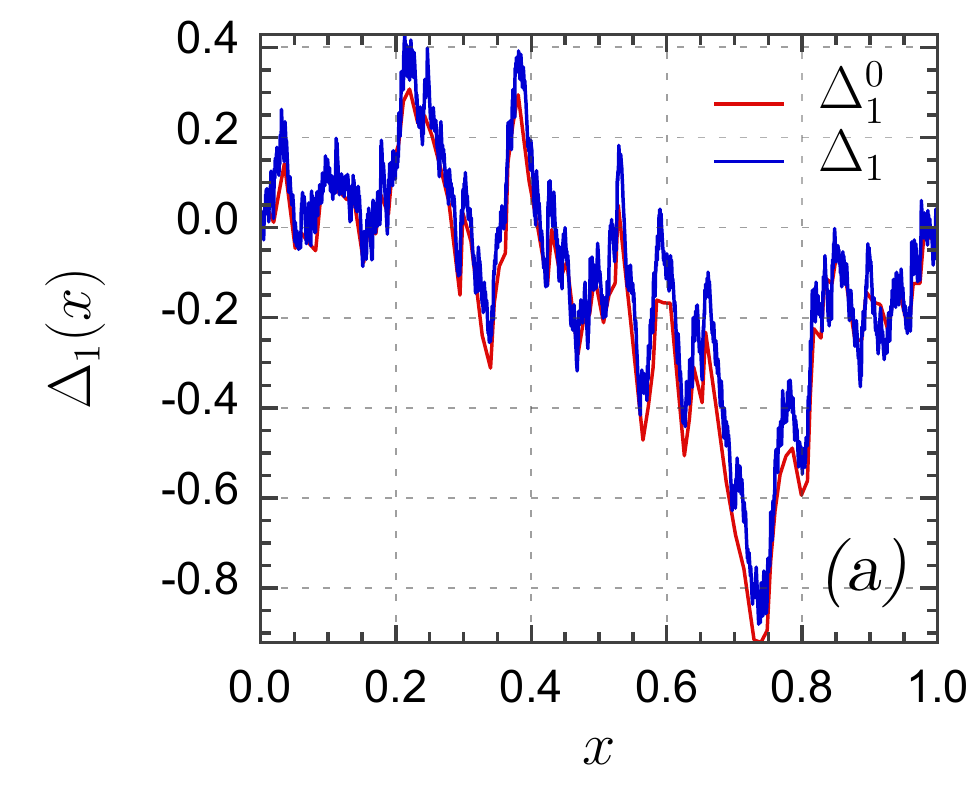}
\includegraphics[width=\figscale\textwidth]{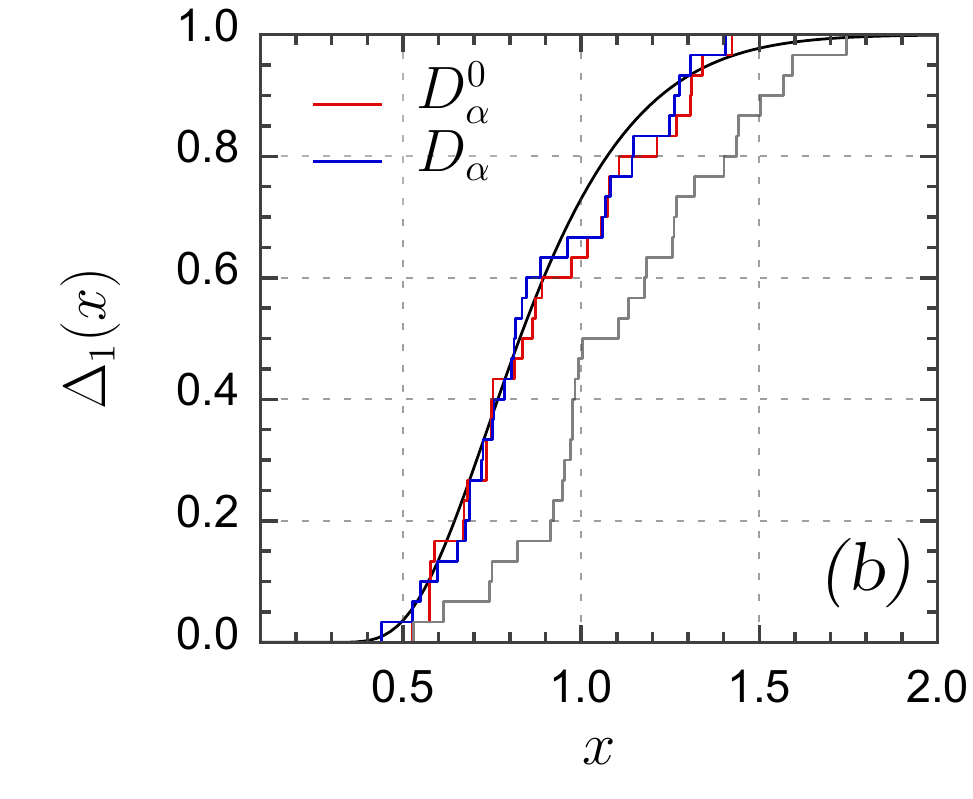}
\caption{(a) Comparison of $\Delta_1^0$ and $\Delta_1$. (b)  Distributions of $D$ for 30 independent samples of 1000 iid uniform variables; the black line is the distribution of Brownian bridge, Eq.~(\ref{PBB}).  The grey line is obtained by adding a small bias to the sampling.}
\label{KSBlockUniform}
\end{center}
\end{figure}

On Fig.~\ref{KSBlockUniform}, we compare the statistics obtained with the uniform law on $[0,1]$: $F(y)=y$.
We build a sequence of 30000 independent values. Then we divide the sequence into 30 blocks.
The first plot illustrates the difference between  $\Delta_1^0(y)=\sqrt N(F_1(y)-y)$ and $\Delta_1(y)=\sqrt N(F_1(y)-\hat F(y))$.
The second plot shows the distribution of $D_\alpha^0=\max_y\Delta_1^0(y)$ and $D_\alpha$.

In turn, the $D_\alpha$'s can be tested against the distribution of maximum of the normalized Brownian bridge. 
We obtain $D=0.688$ and $D^0=0.664$ corresponding to probabilities $P(D>0.688)=0.73$ and $P(D>0.664)=0.77$
(probabilities between 0.1 and 0.9 are satisfying, probabilities close to one indicate not independent sampling).

By contrast, the grey curve in Fig.~\ref{KSBlockUniform}(b),  is obtained by testing a sample of  uniform variables on the interval $[0,1.02]$ against the uniform law on $[0,1]$.
In this case we obtain $D^0=2.59$ corresponding to a probability of $3.0\,10^{-6}$. For the same sampling, the test with the first 10 blocks 
gives respectively $P=0.044$ and with 100 blocks of length 1000 we obtain $P= 2.7\,10^{-87}$.


If the variables are not independent, but not strongly correlated, one can apply the Kolmogorov-Smirnov theorem  to  a subsequence $Y_i^*=Y_{ki}$ where $k$ is of order of the correlation length.

\subsection{Kolmogorov-Smirnov test for Monte Carlo observables}
In our case, the Kolmogorov-Smirnov theorem cannot apply to the process  $\{X_{i},n_{i}\}$  whose distribution function is :
\begin{align}\label{fn}
F_N(x)=\frac{\sum_i \chi(X_i<x)n_i}{\sum_i n_i}.
\end{align}
The equivalent of Eq.~(\ref{fn3}) involves the second moments $E(\chi(X_i<x)n_i^2)$ which cannot be expressed in term of $F(x)$.

\begin{figure}[H]
\begin{center}
\includegraphics[width=\figscale\textwidth]{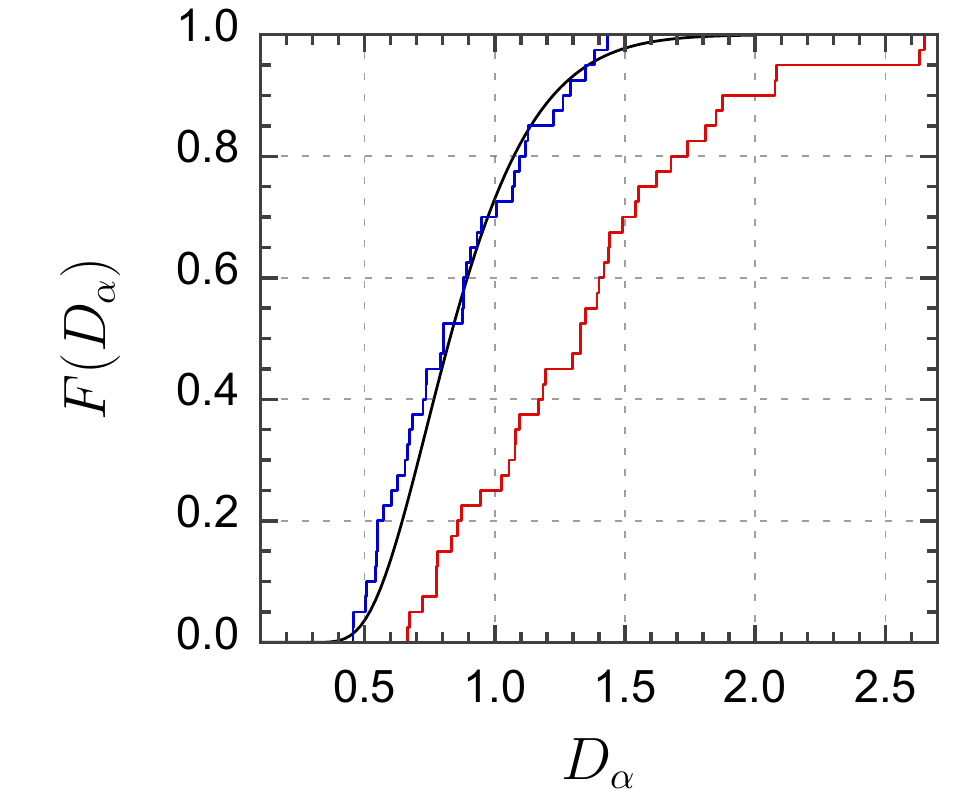}
\caption{Distribution functions of $D_\alpha$. The red curve stands for the original sampling.
The blue curve is obtained with a stride of 15.
}
\label{QMCKS}
\end{center}
\end{figure}

Nevertheless, we can check the law of $X_i$ forgetting the weights $n_i$. Instead of the definition Eq.~(\ref{fn}), we use:
\begin{align}\label{gn}
G_N(x)=\frac 1N\sum_i \chi(X_i<x).
\end{align}
As above, we have $p$ samples of length $N$ obtained by $p$ equivalent QMC runs. 
For each sample we compute the distribution function $G_\alpha(y)$ and the distribution  $\hat G(y)$ for the union of the samples, Eq.~(\ref{fn}).
Then we set:
\begin{align}
D_\alpha&={\sqrt N}\sup_y \left|G_\alpha(y)-\hat G(y)\right|,
\end{align}
Thus, for independent variables, $D_\alpha$ has the law of the maximum absolute value of the Brownian bridge.
Fig.~\ref{QMCKS} shows the distribution of $D_\alpha$ (red curve). The black curve represents the Brownian bridge.
We have 40 samples of about 20000 trials.

The KS test gives a probability of $7.9\, 10^{-10}$ confirming that our samples do not correspond to a sampling of independent variables.
The blue curve is obtained with smaller samples retaining one trial for 15 original trials. Indeed, we have $C_{15}/C_0=0.01$ while $C_1/C_0=0.3$, see Eq.~(\ref{Ckdef}).
The probability given by the KS test is 0.53 for the blue curve. The new  samples can be considered as made of independent variables and  sharing the same law.

In Fig.~\ref{QMCKS}, the vertical steps of the blue curve are equal to $1/p$ and the vertical distance to the black curve should be of order $1/\sqrt p$.
A Ugly Duckling in the sampling results in a single large distance $D_\alpha$, inducing a large horizontal step of height still $1/p$ at the top of figure,
but such a step has no significant effect on this KS test.
One can detect these inconsistencies by testing the maximum $c$ of the $D_\alpha$ using 
$$P(\max_\alpha D_\alpha>c)=1- P(D<c)^k$$ where $k$ is the number of samples. In our case we get $c=2.65$ (resp. 1.43) and $P=6\,10^{-5}$ (resp. 0.26).\\

Thus the data providing the red curve of Fig.~\ref{QMCKS} is rejected with both tests.

\section{Efficiency of VMC}
The efficiency of the VMC may be defined as the accuracy obtained with a given CPU time.
The standard deviation of the results is given by $\seff/\sqrt N$ where $N$ is the number of retained energies.
The CPU time is proportional to the number of steps $M=NE(n)=N/\ra$. Thus the efficiency of the VMC may be measured with the dimensionless parameter 
\begin{align}
\Q=\frac {\sigma \sqrt \ra}\seff
\end{align}
where $\sigma$ is the standard deviation given by the distribution of the energy;
i.e. the standard deviation of the VMC for $M$ trials rewrites:
\begin{align}
\frac{\seff}{\sqrt N}=\frac\sigma{\Q\sqrt M}.
\end{align}
Notice that $\sigma$ is a lower bound for $\seff$, therefore $\Q\le1$.
\subsection{Experimental results}
\begin{figure}[H]
\begin{center}
\includegraphics[width=\figscale\textwidth]{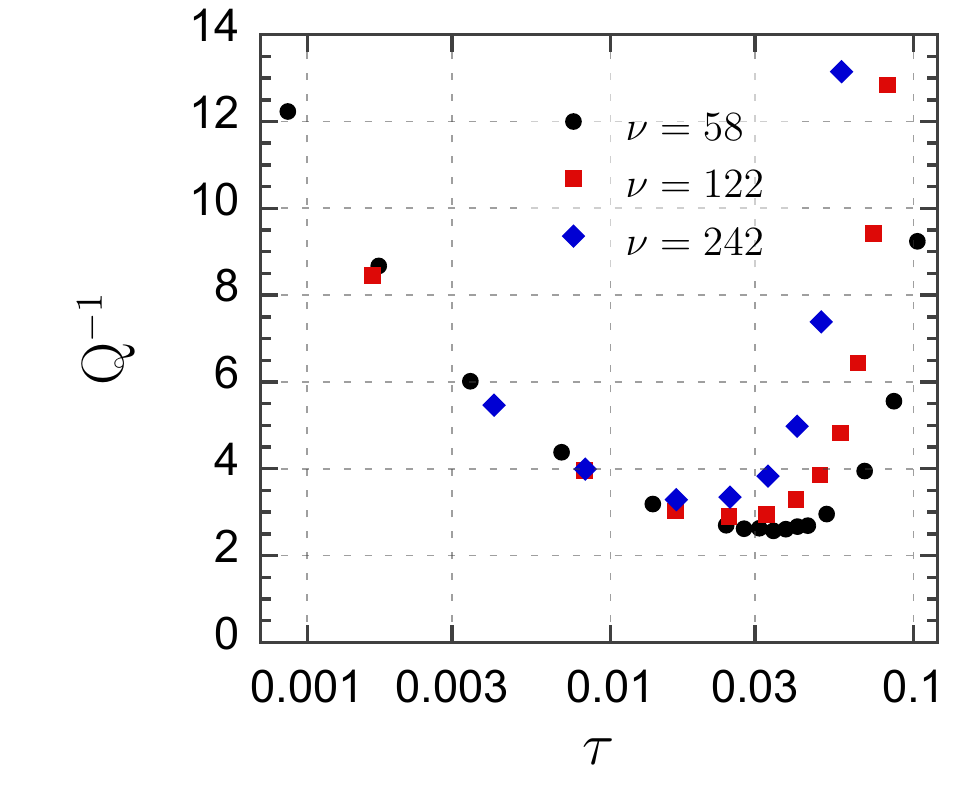}
\includegraphics[width=\figscale\textwidth]{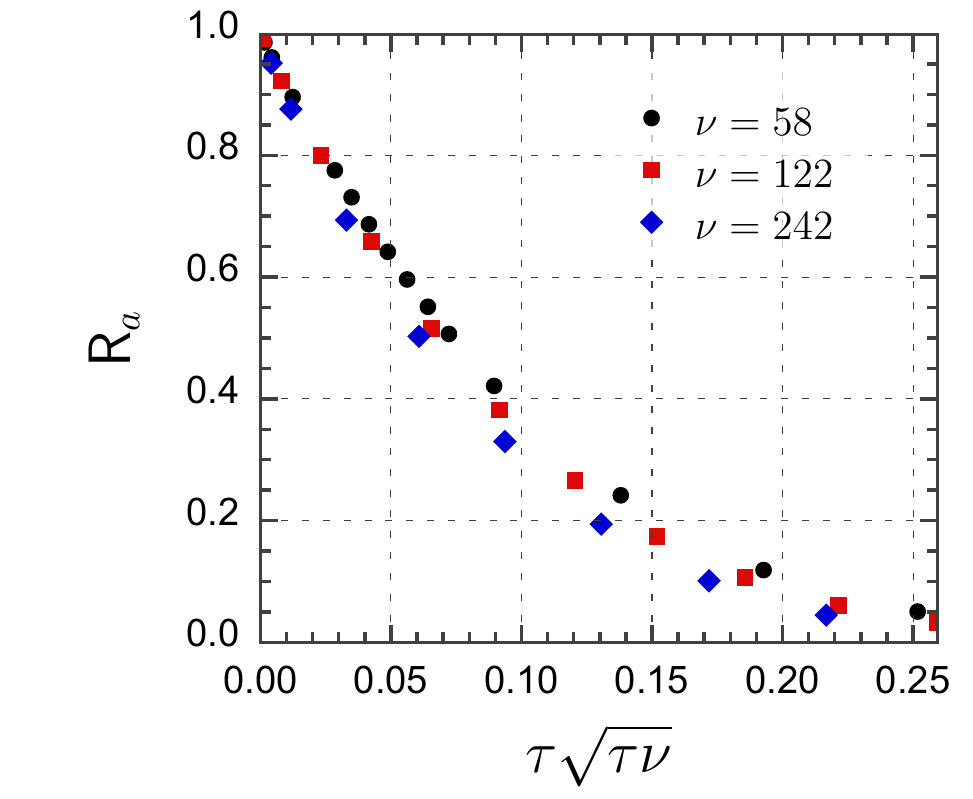}
\caption{Inverse efficiency, $\Q^{-1}$, and acceptance probability, $\ra$, for different values of the Monte Carlo diffusion time step, $\tau$,
and the dimension of configuration space, $\nu$.
}
\label{sigmaeff}
\end{center}
\end{figure}
Fig.~\ref{sigmaeff} gives the behavior of $\Q$ (resp. $\ra$) as a function of $\tau$ (resp. $\tau\sqrt{\tau\nu}$), the time step of the discretized diffusion process. 
We see that for small $\tau$, $\Q$ is a function of $\tau$ only ($\tau$ is normalized in the sense that by Eq.~(\ref{kernel}) the mean distance between particles does not depend on $\nu$).
The behavior of $\ra$ as a function of $\tau\sqrt{\tau\nu}$ is more questionable since the range of $\nu$ is rather limited.
Nevertheless, as $\nu$ increases, smaller $\tau$ must be chosen to obtain similar behavior. The behavior of $\Q$ for intermediate values of $\tau$
(Fig.\ref{sigmaeff}) enforces this relationship between VMC and $\nu$.
As it is usually claimed  the minimal $\Q$ is for $\ra\approx 0.5$.

The scaling law of $\ra$ may be understood as follows. As $\tau$ goes to zero, we have  at the leading order:
\begin{align}
\label{Qapprox}
A(R',R)=\min\left(1,e^{\frac\tau2 \left(G(R)^2-G(R')^2\right)}\right)
\end{align}
Let  $H$ be the gradient of $G$ (i.e. the Hessian of $\frac 12\ln P$) then
\begin{align}
G(R)^2-G(R')^2\approx 2(R-R')\cdot HG.
\end{align}
Now $R-R'\approx\sqrt\tau\eta_\nu$ where $\eta_\nu$ is a vector of $\nu$ normalized Gaussian variables, thus: 
\begin{align}
A(R',R)\approx\min\left(1,e^{\tau \sqrt\tau \|HG\|\eta}\right) .
\end{align}
where $\eta$ is a normalized Gaussian variable. Therefore the acceptance at $R$ is
\begin{align}
\ra(R)&\approx1-\frac{\tau \sqrt\tau}{\sqrt{2\pi} }\|HG\|.
\end{align}
Assuming that $H$ is bounded, we have $\|HG\|^2\propto\nu$ and
\begin{align}
\label{Qapprox2}
\ra&\approx1-\frac{\tau \sqrt{\tau\nu}}{\sqrt{2\pi} }c
\end{align}
Be aware that $c$ in Eq.~(\ref{Qapprox2}) depends on $P$ and thus on $\nu$ (it is not simple to build equivalent models with different $\nu$).
Figure~\ref{sigmaeff} is built from different wave functions, $\Psi$, representing the electron gas at different number of particles but at the same density.

\begin{figure}[H]
\begin{center}
\includegraphics[width=\figscale\textwidth]{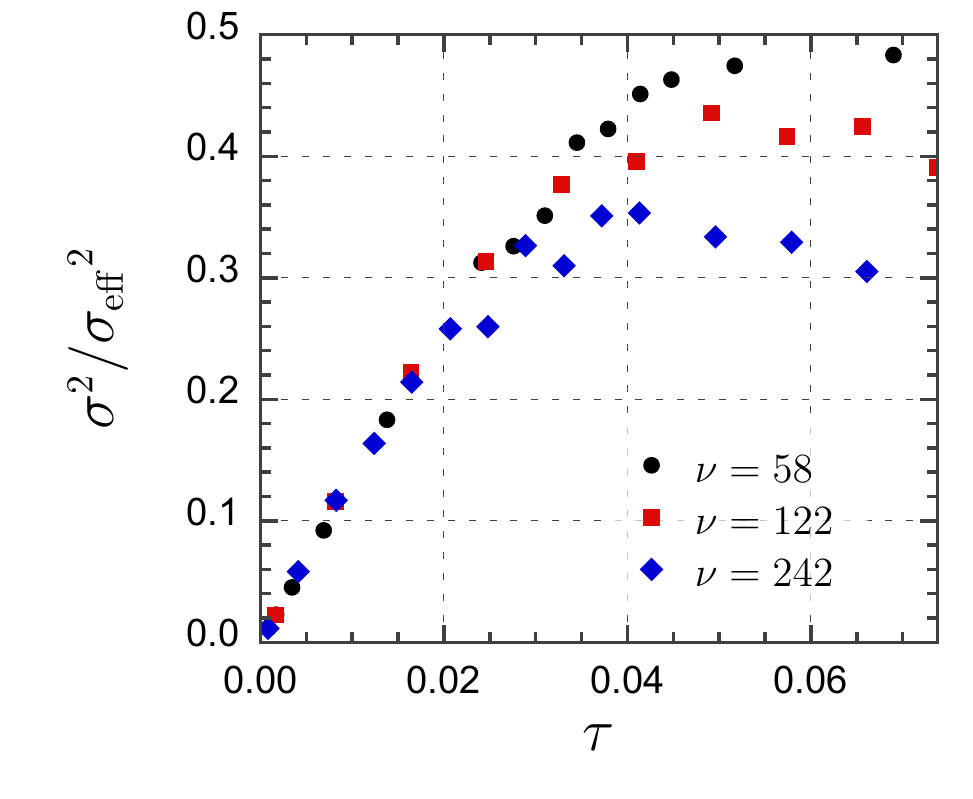}
\caption{$\sigma^2/\seff^2$ as a function of the time discretization. Different patterns correspond to different values of $\nu$ ($\sigma^2=0.04, 0.02, 0.01$).
}
\label{sigmaratio}
\end{center}
\end{figure}
Figure~\ref{sigmaratio} reveals the linear behavior of $\sigma^2/\seff^2$ for small values of $\tau$. 
Indeed, as $\tau$ goes to zero, we expect that the the discrete process converges to the continuous process, Eq.~(\ref{wiener});
in particular $k$ steps at time discretization $\tau$ are equivalent to one step at time discretization $k\tau$. Thus  the correlation length scales as  $1/\tau$ and
 $\seff^2/C_0$ behaves  like $1/\tau$. At the same time, $C_0$ goes to $\sigma^2$ the variance of energy since $E(n)=1/\ra$ goes to 1.
 Therefore for small $\tau$:
\begin{align}
 \Q^{-1}\approx\frac \seff\sigma\propto \frac 1{\sqrt\tau}
\end{align}
On the other hand, as $\ra$ goes to zero the trials become independent; thus $\seff^2$ goes to $C_0$ and Eq.~(\ref{corrN}) gives 
\begin{align}
\label{corrN0}
C_0&\longrightarrow\frac 1{ E(n)(\sum_i n_i)} \sum_{i} n_i^2\tilde X_i\propto \sigma^2
\end{align}
Thus for small $\ra$:
\begin{align}
\Q^{-1}\approx\frac \seff{\sigma\sqrt{\ra}}&\propto \frac 1{\sqrt{\ra}}
\end{align}
\subsection{Implementaion details}
In practice, a raw implementation of QMC leads to blocking configurations as soon as $\tau$ is not very small.
Indeed, the probability $P(R)$ comes from an antisymmetric wave function and thus vanishes at least at $r_i=r_j$.
If $P(R)$ vanishes then $G(R)$ diverges and this results in very large values of $n_i$.
As $\tau$ increases, $|R-R'|$ increases and the nature of the process changes: 
Eq.~(\ref{Qapprox}) is a second order approximation which may be no more 
relevant.
In Eq.\ref{Q}, the factor $\frac{T(R|R')}{T(R'|R)}$ may be very small.
Setting $\eta=R'-R-\tau G(R)$, we have:
\begin{align}
\nonumber
\frac{T(R|R')}{T(R'|R)}&=\exp\left(\frac1{2\tau}\eta^2-\frac1{2\tau}(\eta+\tau G(R)+\tau G(R'))^2\right)\\
\label{expo}
&=\exp\left(-\eta(G(R)+G(R'))-\frac \tau 2(G(R)+G(R'))^2\right)
\end{align}
Here, $\eta$ is a vector  of $\nu$ Gausssian variables of variance $\tau$, so $\eta G(R)$ is a Gaussian variable of variance $\tau G(R)^2$.
Therefore, if $G(R)^2$ is large, the exponent in Eq.~(\ref{expo}) is of order $-\tau G(R)^2$, and, in our simulations, we have to ensure that drift $\tau G(R)$ is bounded, otherwise the exponent may be very large leading to blocking situations. 

To avoid these situations, we have rescaled $G$ to ensure that $\tau G^2$  is always bounded by $m_G\nu$.
\begin{figure}[H]
\begin{center}
\includegraphics[width=\figscale\textwidth]{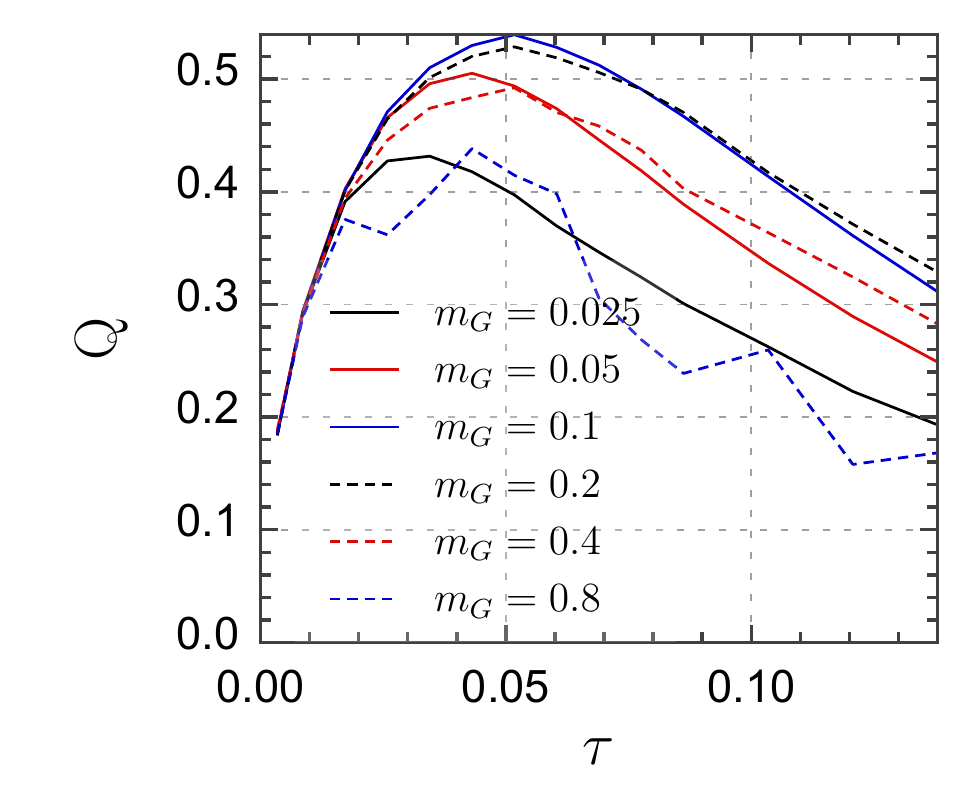}
\caption{Efficiency $\Q$ as a function of $\tau$ for for different values of $m_G$ ($\nu=58$ and $r_s=1$).
}
\label{maxD}
\end{center}
\end{figure}
From Fig.~\ref{maxD} we see that $m_G\approx 0.05$-$0.4$ gives the best results. For $m_G=0.8$, the fluctuations of $Q$ becomes important: blocking states occur and the statistics  
are no more  usable.

This has been checked also for $r_s$ (resp. $\nu$) up to $35$ (resp. $242$). For larger values of $\nu$ the value of $m_G$ is less critical: probably the law of large numbers
makes that the relative fluctuations of $G^2$ decrease.

\subsection{Asymptotic law of $P$}
For 20000 trials, $P=|\Psi|^2$ has been found in a range 1-$10^{13}$. This may seem surprising; in fact, this behavior is banal 
and appears in standard models in statistical physics.
As an example, let us consider the momentum distribution of the ideal gas:  
\begin{align}
P(p_1,\ldots ,p_\nu)d^\nu p&=\frac{e^{-\frac{\sum_i p_i^2}{2}}}{\sqrt{2\pi}^\nu}d^\nu p
\end{align}
then setting $\phi=\sum_i p_i^2/\nu$, the image (or push-forward) measure of $d^\nu p$ is $V(\phi)d\phi$ where $V(\phi)\propto \phi^{\frac{\nu-2}2} $.
Thus the the image measure of $Pd^\nu p$ is proportional to:
\begin{align}
\rho(\phi)d\phi=\phi^{\frac{\nu-2}2} e^{-\frac{\nu \phi}{2}}d\phi
\end{align}
Then as $\nu$ goes to $\infty$, with probability one $\phi$ is the maximum $\phi_0$ of $\ln(\phi)-\phi$,
i.e. $\phi_0=1$, and around $\phi_0$
\begin{align}
\rho(\phi)d\phi\approx \exp\left(-\nu\frac {(\phi -\phi_0)^2}4\right)d\phi
\end{align}
That is $\phi$ has fluctuations of order $1/\sqrt\nu$  and while $\rho(\phi)$ is almost constant, $V(\phi)$ and $e^{-\frac{\nu \phi}{2}}$
have large fluctuations of order $e^{\pm C\sqrt\nu}$.

\begin{figure}[H]
\begin{center}
\includegraphics[width=\figscale\textwidth]{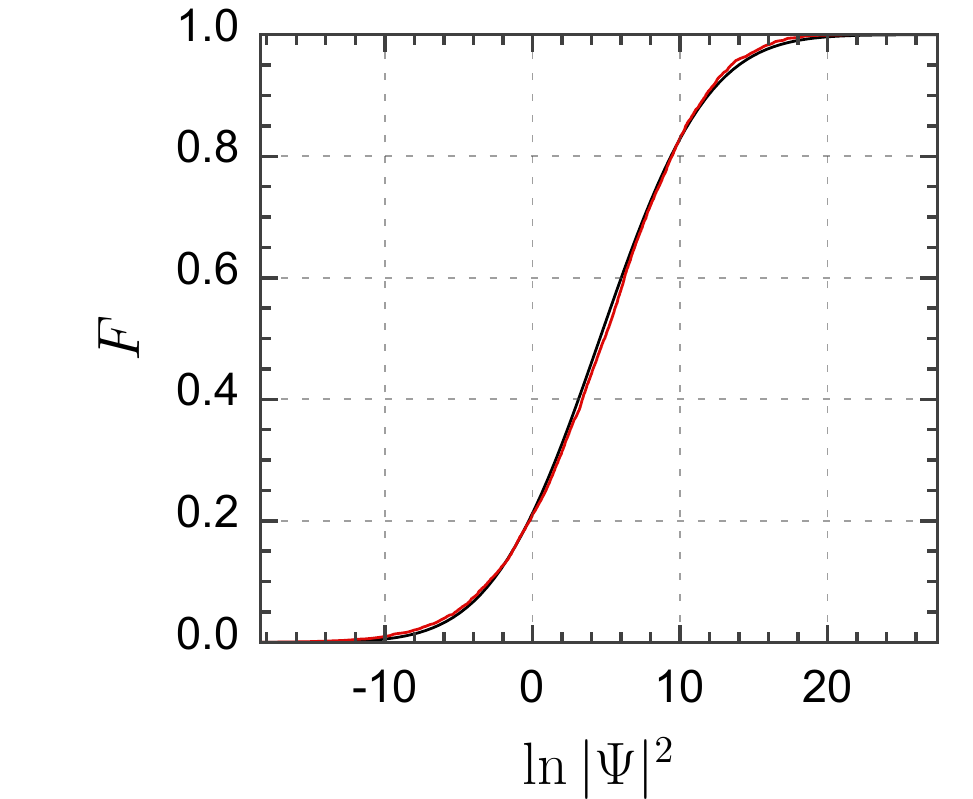}
\caption{Repartition function of $\ln |\Psi|^2$ ($\nu=122$ and $r_s=10$).
Black line is the normal law; red line is obtained with 20000 trials.
}
\label{psi2}
\end{center}
\end{figure}

In our case, $\Psi$ is not normalized, thus the mean of $\ln |\Psi|^2$ (proportional to $\nu$) does not make sense but the Gaussian law is clear with a standard deviation $\sigma/\sqrt \nu=0.52$ (Fig.\ref{psi2}).

The variance of  $\ln |\Psi|^2$ should depend on the choice of $\Psi$ but the Gaussian law is expected in any case for large $\nu$; 
indeed, the Jastrow part of $\Psi$ is usually a product of $\nu^2$ factors and the Slater part is a determinant which is also, in the thermodynamics 
limit, the exponential of some extensive quantity.

\section{Conclusion}
In this paper, we have reviewed well known results  of statistics applied to Monte Carlo calculations. We have
provided effective algorithms to compute the accuracy and to check the equilibration of Monte Carlo simulations.
In order to optimize and understand the limitations of standard variational Monte Carlo sampling, we have further
 described general scaling laws of the discrete time approximation of the diffusion process with examples on the
 homogeneous two dimensional electron gas.

\end{document}